\documentclass[pre,aps,floats,twocolumn,superscriptaddress,floatfix]{revtex4}
\usepackage{graphicx,amssymb,amsmath,ifthen,rotating}
\usepackage[active]{srcltx}
\usepackage{tikz}
\usepackage{verbatim}
\begin{document}
\title{Jammed disks of two sizes and weights in a channel:
Alternating sequences}  
  \author{Dan Liu}
\affiliation{
  Department of Physics,
  University of Hartford,
  West Hartford, CT 06117, USA}
\author{Gerhard M{\"{u}}ller}
\affiliation{
  Department of Physics,
  University of Rhode Island,
  Kingston RI 02881, USA}
\begin{abstract}
Disks of two sizes and weights in alternating sequence are confined to a long and narrow channel. 
The axis of the channel is horizontal and its plane vertical.
The channel is closed off by pistons that freeze jammed microstates out of loose disk configurations subject to moderate pressure, gravity, and random agitations.
Disk sizes and channel width are such that under jamming no disk remains loose and all disks touch one wall.
We present exact results for the characterization of jammed macrostates including volume and entropy.
The rigorous analysis divides the disk sequences of jammed microstates into overlapping tiles from which we construct a small number of species of statistically interacting particles. 
Jammed macrostates depend on dimensionless control parameters inferred from ratios between measures of expansion work against the pistons, gravitational potential energy, and intensity of random agitations.
These control parameters enter the configurational statistics via the activation energies prior to jamming of the particles.
The range of disk weights naturally divides into regimes where qualitatively different features come into play. 
We sketch a path toward generalizations that include random sequences under a modified jamming protocol.
\end{abstract}
\maketitle
%
\section{Introduction}\label{sec:intro}
%
The physics of jammed granular matter is dominated by experimental studies \cite{KFL+95, MJS00, TPC+01, LDDB06, SNRS07, MSLB07, GBOS09, ZXC+09, ZMSB10, NJVB11, ENW12, PD13, BKD19} and computational studies \cite{OLLN01, OLLN02, XBO05, ZM05, UKW05, SDST06, DST06, GBO06, OSN06, SVE+07, SWM08, Head09, DW09, JCM+10, CCC10, JM10, SOS11, XFL11, OH11, CPNC11, PCC12, CCPZ12, HB12, LFS+18} more than is the case in many other areas of condensed matter physics. 
The advanced computational power now available for simulations of granular matter subject to diverse protocols of athermal agitations and jamming has efficiently aided the interpretation of existing measurements and has inspired alternative experimental probes.
The wealth of phenomena, too rich in scope to be listed here, has been reviewed from a range of perspectives \cite{JN96, DeGe99, MBE05, Meht10, TS10, Xu11, BMHM18, BB19}.

Theoretical approaches operating under the umbrella of \emph{configurational statistics} \cite{EO89, AE89, ME94, EM94, EG99, Edwa08, BE09, WSWM12} have carved out comparably small corners in this wide field of research which permit an exact analysis or systematic approximations of some scenarios by specifically designed methodologies \cite{BB02, SST03, BE03, BSWM08, BSWM10, DJM10, SWJM10, MSJ+10, DeMc11}. 
One such corner involves jammed matter in narrow channels \cite{LM07, LM10, BS06, AB09, BA11, YAB12, AYB13, janac1, janac2}, where the diversity of jammed configurations is limited by \emph{caging effects} \cite{LFS+18}. 
Nevertheless, the constrained quarters of granular matter in narrow channels leave ample room for complex phenomena and promising routes for the further development of methods of exact analysis. 

The research reported in this work builds on two previous studies \cite{janac1,janac2}, in which jammed granular matter in a narrow channel is represented by configurations of statistically interacting particles and amenable to exact analysis by methods of statistical mechanics developed in different contexts \cite{Hald91a, Wu94, Isak94, Anghel, NA14, LVP+08, copic, picnnn, pichs} and adapted to configurational statistics.
Their focus was on the ordering tendencies of jammed disks of one size and weight, confined to a narrow channel and subject to combinations of gravity, friction, and centrifugation.

Here we consider configurations of jammed disks with two different sizes and two different weights, which adds diversity to the observed phenomena.
More importantly, this generalization is a way station toward the investigation of random configurations of disks with different sizes and weights. 
Exact results for a random sequence of jammed disks with two sizes and weights thus move well within reach in the wake of this study.

The key step that makes this progress possible is a twofold shift in the entities subjected to configurational statistics, (i) from disks to tiles and (ii) from tiles to statistically interacting particles.
Step (i) was already taken in Refs.~\cite{BS06, AB09, BA11}. 
Step (ii)  was developed in Refs.~\cite{janac1, janac2} for disks of one size and is here generalized to situations involving disks of two sizes.
The analysis is laid out in a way that facilitates further generalizations to random sequences of such disks.

The methodology presented here for jammed granular matter is far from common and it is not yet clear where it will reach its limitations.
It will remain promising for as long as it produces new exact results of interest. 
Interestingly, the same methodology also yields new exact results in applications to quantum gases \cite{PMK07}, to quantum spin chains embedded in solids \cite{LMK09}, and to biological matter including polypeptides \cite{cohetra} and DNA \cite{mct1}.

The configurational statistics of disks with two different sizes and/or weights jammed in a channel includes ingredients related to geometry (Sec.~\ref{sec:geom}), energetics (Sec.~\ref{sec:ener}), combinatorics (Sec.~\ref{sec:comb}), and statistical mechanics (Sec.~\ref{sec:stat}) to be worked out up front and then assembled.
We discuss each ingredient in the context of the simplest application, where the weights of small and large disks are in balance (Sec.~\ref{sec:we-in-ba}), and then use them again for applications, where size and weight produce more complex patterns (Secs.~\ref{sec:he-la-di} and \ref{sec:li-la-di}). 

In the absence of gravitational effects, the statistical mechanics can be carried out with just two species of particles instead of five (Sec.~\ref{sec:ze-gra}). 
We conclude by pointing the way toward a generalization of this work which includes jammed macrostates of random sequences of disks with two sizes and weights. The modified jamming protocol permits disks to move past each other in a widened channel during random agitations (Sec.~\ref{sec:outl}).

%
\section{Geometry}\label{sec:geom}
%
Disks of two sizes with diameters $\sigma_\mathrm{L}\geq\sigma_\mathrm{S}$ in alternating sequence are being jammed in a channel of width $H$.
Every jammed disk has three points of contact, either with an adjacent disk or with a wall. 
The constraints, 
\begin{subequations}\label{eq:1} 
\begin{align}\label{eq:1a} 
& \frac{1}{4}<\frac{\sigma_\mathrm{S}}{\sigma_\mathrm{L}}\leq 1, 
\\ \label{eq:1b} 
& 1<\frac{H}{\sigma_\mathrm{L}}<
\frac{1}{2}\left[1+\frac{\sigma_\mathrm{S}}{\sigma_\mathrm{L}} +
\sqrt{\left(1+\frac{\sigma_\mathrm{S}}{\sigma_\mathrm{L}}\right)^2-1}\;\right],
\end{align}
\end{subequations}
guarantee that the disk sequence remains invariant before jamming at constant $H$, that all jammed disks have wall contact, and that no loose disks exist under jamming.
In the one-size limit $\sigma_\mathrm{S}/\sigma_\mathrm{L}\to1$, condition (\ref{eq:1b}) reduces to the condition, $1<H/\sigma<1+\sqrt{3/4}$, familiar from Ref.~\cite{janac1}.

All jammed microstates can be assembled from 8 tiles composed of two adjacent disks similar to domino pieces [Table~\ref{tab:t1}], with a one-disk overlap.
Adding a tile to an already existing string must satisfy two successor rules:
The tile added must (i) match the pattern regarding size and position of the overlapping disk and (ii) maintain mechanical stability under jamming forces.
Each tile has one of two distinct volumes [Table ~\ref{tab:t2}].

\begin{table}[htb]
  \caption{Distinct tiles that constitute jammed microstates of alternating disk sequences subject to the constraints (\ref{eq:1}). Mechanical stability rule: \textsf{v} must be followed by \textsf{w} or 2 etc. Motifs pertain to $\sigma_\mathrm{L}=2$, $\sigma_\mathrm{S}=1.4$, $H=2.5$.}\label{tab:t1}
\begin{center}
\begin{tabular}{cccc|cccc} \hline\hline
motif & ~ID~ & rule & vol.~ & ~motif~ & ~ID~ & rule & vol.  \\ \hline \rule[-2mm]{0mm}{8mm}
\begin{tikzpicture} [scale=0.2]
\draw (0.0,0.0) -- (3.2,0.0) -- (3.2,2.5) -- (0.0,2.5) -- (0,0);
\filldraw [fill=gray, draw=black] (1,1) circle (1.0);
\filldraw [fill=gray, draw=black] (2.5,1.8) circle (0.7);
\end{tikzpicture}
& \textsf{v} & \textsf{w}, 2 & $V_\mathrm{c}$ &   
\begin{tikzpicture} [scale=0.2]
\draw (0.0,0.0) -- (3.37,0.0) -- (3.37,2.5) -- (0.0,2.5) -- (0,0);
\filldraw [fill=gray, draw=black] (0.7,0.7) circle (0.7);
\filldraw [fill=gray, draw=black] (2.37,1.0) circle (1.0);
\end{tikzpicture}
& 3 & \textsf{v} & $V_\mathrm{f}$  \\ \rule[-2mm]{0mm}{6mm}

\begin{tikzpicture} [scale=0.2]
\draw (0.0,0.0) -- (3.2,0.0) -- (3.2,2.5) -- (0.0,2.5) -- (0,0);
\filldraw [fill=gray, draw=black] (0.7,1.8) circle (0.7);
\filldraw [fill=gray, draw=black] (2.2,1.0) circle (1.0);
\end{tikzpicture}
& \textsf{w} & \textsf{v}, 1 & $V_\mathrm{c}$ &  
\begin{tikzpicture} [scale=0.2]
\draw (0.0,0.0) -- (3.2,0.0) -- (3.2,2.5) -- (0.0,2.5) -- (0,0);
\filldraw [fill=gray, draw=black] (0.7,0.7) circle (0.7);
\filldraw [fill=gray, draw=black] (2.2,1.5) circle (1.0);
\end{tikzpicture}
& 4 & 5, 6 & $V_\mathrm{c}$  \\ \rule[-2mm]{0mm}{6mm}

\begin{tikzpicture} [scale=0.2]
\draw (0.0,0.0) -- (3.37,0.0) -- (3.37,2.5) -- (0.0,2.5) -- (0,0);
\filldraw [fill=gray, draw=black] (1.0,1.0) circle (1.0);
\filldraw [fill=gray, draw=black] (2.67,0.7) circle (0.7);
\end{tikzpicture}
& 1 & 3, 4 & $V_\mathrm{f}$ & 
\begin{tikzpicture} [scale=0.2]
\draw (0.0,0.0) -- (3.2,0.0) -- (3.2,2.5) -- (0.0,2.5) -- (0,0);
\filldraw [fill=gray, draw=black] (1,1.5) circle (1.0);
\filldraw [fill=gray, draw=black] (2.5,0.7) circle (0.7);
\end{tikzpicture}
& 5 & 3, 4 & $V_\mathrm{c}$  \\ \rule[-2mm]{0mm}{6mm}

\begin{tikzpicture} [scale=0.2]
\draw (0.0,0.0) -- (3.37,0.0) -- (3.37,2.5) -- (0.0,2.5) -- (0,0);
\filldraw [fill=gray, draw=black] (0.7,1.8) circle (0.7);
\filldraw [fill=gray, draw=black] (2.37,1.5) circle (1.0);
\end{tikzpicture}
& 2 & 5 & $V_\mathrm{f}$ & 
\begin{tikzpicture} [scale=0.2]
\draw (0.0,0.0) -- (3.37,0.0) -- (3.37,2.5) -- (0.0,2.5) -- (0,0);
\filldraw [fill=gray, draw=black] (1.0,1.5) circle (1.0);
\filldraw [fill=gray, draw=black] (2.67,1.8) circle (0.7);
\end{tikzpicture}
& 6 & \textsf{w}, 2 & $V_\mathrm{f}$  \\ \hline\hline
\end{tabular}
\end{center}
\end{table}

\begin{table}[htb]
  \caption{Volume of tiles (with unit cross section).  
  The numerical values are for $\sigma_\mathrm{L}=2$, $\sigma_\mathrm{S}=1.4$, $H=2.5$.}\label{tab:t2}
\begin{center}
\begin{tabular}{l|l|l} \hline\hline
 & ~vol. & ~num.  \\ \hline \rule[-2mm]{0mm}{6mm}
$V_\mathrm{c}$~ & ~$\frac{1}{2}(\sigma_\mathrm{L}+\sigma_\mathrm{S})
+\sqrt{H(\sigma_\mathrm{L}+\sigma_\mathrm{S}-H)}$~~ & ~3.2 
\\ \rule[-2mm]{0mm}{6mm}
$V_\mathrm{f}$~ & ~$\frac{1}{2}(\sigma_\mathrm{L}+\sigma_\mathrm{S})
+\sqrt{\sigma_\mathrm{L}\sigma_\mathrm{S}}$ & ~3.373 
\\ \hline\hline
\end{tabular}
\end{center}
\end{table}

Next we choose a convenient jammed reference state and generate other jammed microstates via the activation of quasiparticles which modify the reference state.

%
\section{Energetics}\label{sec:ener}
%
With no loss of generality regarding macrostates, we assume to have $N$ (large or small) pairs of disks in the channel with the first (large) disk and the last (small) disk fixed in the (vertical) positions of tile \textsf{v}.
Under these assumptions the microstate of minimum volume is unique.
It is composed of an alternating sequence of just two tiles:
\begin{equation}\label{eq:2} 
\mathrm{pv}=\mathsf{vwvw\cdots v}.
\end{equation}
We declare it to be the pseudo-vacuum for statistically interacting particles in this application.
All other (jammed) microstates can be generated by the activation of quasi-particles from this reference state. 
We have identified $M=5$ species of particles that serve this purpose [Table~\ref{tab:t3}].
Adopting the taxonomy of \cite{copic} we distinguish between the categories of hosts and tags.

\begin{table}[htb]
  \caption{Five species of quasi-particles. The hosts $m=1,\ldots,4$ modify the pseudo-vacuum (\ref{eq:2}) whereas the tag $m=5$ modifies any one of the hosts. The motifs shown are for $\sigma_\mathrm{L}=2$, $\sigma_\mathrm{S}=1.4$, $H=2.5$. The ID lists the sequence of tiles that make the particle. The excess volume is $\Delta V_m$ and the activation energy $\epsilon_m$.}\label{tab:t3}
\begin{center}
\begin{tabular}{llllll} \hline\hline
motif & ID & $m$ & cat. & $\Delta V_m$ & $\epsilon_m$  \\ \hline \rule[-2mm]{0mm}{8mm}
\begin{tikzpicture} [scale=0.2]
\draw (0.0,0.0) -- (5.34,0.0) -- (5.34,2.5) -- (0.0,2.5) -- (0,0);
\filldraw [fill=gray, draw=black] (1.0,1.0) circle (1.0);
\filldraw [fill=gray, draw=black] (2.67,0.7) circle (0.7);
\filldraw [fill=gray, draw=black] (4.34,1.0) circle (1.0);
\end{tikzpicture}
& \textsf{13} & 1 & host & $2V_\mathrm{t}$ & $2pV_\mathrm{t}-\gamma_S$ 
\\ \rule[-2mm]{0mm}{6mm}
\begin{tikzpicture} [scale=0.2]
\draw (0.0,0.0) -- (6.54,0.0) -- (6.54,2.5) -- (0.0,2.5) -- (0,0);
\filldraw [fill=gray, draw=black] (1.0,1.0) circle (1.0);
\filldraw [fill=gray, draw=black] (2.67,0.7) circle (0.7);
\filldraw [fill=gray, draw=black] (4.17,1.5) circle (1.0);
\filldraw [fill=gray, draw=black] (5.84,1.8) circle (0.7);
\end{tikzpicture}
& \textsf{146} & 2 & host & $2V_\mathrm{t}$ & $2pV_\mathrm{t}-\gamma_\mathrm{S}+\gamma_\mathrm{L}$ 
\\ \rule[-2mm]{0mm}{6mm}
\begin{tikzpicture} [scale=0.2]
\draw (0.0,0.0) -- (6.54,0.0) -- (6.54,2.5) -- (0.0,2.5) -- (0,0);
\filldraw [fill=gray, draw=black] (0.7,1.8) circle (0.7);
\filldraw [fill=gray, draw=black] (2.37,1.5) circle (1.0);
\filldraw [fill=gray, draw=black] (3.87,0.7) circle (0.7);
\filldraw [fill=gray, draw=black] (5.54,1.0) circle (1.0);
\end{tikzpicture}
& \textsf{253} & 3 & host & $2V_\mathrm{t}$ & $2pV_\mathrm{t}-\gamma_\mathrm{S}+\gamma_\mathrm{L}$
\\ \rule[-2mm]{0mm}{6mm}
\begin{tikzpicture} [scale=0.2]
\draw (0.0,0.0) -- (7.74,0.0) -- (7.74,2.5) -- (0.0,2.5) -- (0,0);
\filldraw [fill=gray, draw=black] (0.7,1.8) circle (0.7);
\filldraw [fill=gray, draw=black] (2.37,1.5) circle (1.0);
\filldraw [fill=gray, draw=black] (3.87,0.7) circle (0.7);
\filldraw [fill=gray, draw=black] (5.37,1.5) circle (1.0);
\filldraw [fill=gray, draw=black] (7.04,1.8) circle (0.7);
\end{tikzpicture}
& \textsf{2546} & 4 & host & $2V_\mathrm{t}$ & $2pV_\mathrm{t}-\gamma_\mathrm{S}+2\gamma_\mathrm{L}$
\\ \rule[-2mm]{0mm}{6mm}
\begin{tikzpicture} [scale=0.2]
\draw (0.0,0.0) -- (4.4,0.0) -- (4.4,2.5) -- (0.0,2.5) -- (0,0);
\filldraw [fill=gray, draw=black] (0.7,0.7) circle (0.7);
\filldraw [fill=gray, draw=black] (2.2,1.5) circle (1.0);
\filldraw [fill=gray, draw=black] (3.7,0.7) circle (0.7);
\end{tikzpicture}
& \textsf{45} & 5 & tag & $0$ & $\gamma_\mathrm{L}-\gamma_\mathrm{S}$
\\ \hline\hline
\end{tabular}
\end{center}
\end{table}

Each particle consists of overlapping tiles.
Particles from species $m=1,\ldots,4$ can be placed directly into the pseudo-vacuum,
meaning that it is possible to add a tile \textsf{v} or \textsf{w} to the left or to the right as follows:
\begin{equation}\label{eq:11} 
\mathsf{w13v},\quad \mathsf{w146w},\quad \mathsf{v253v},\quad \mathsf{v2546w}.
\end{equation}
Particles from species $m=5$ are parasitic in the sense that they can only be placed inside a particle from species $m=1,\ldots,4$, at exactly one position and with one disk overlapping:
\begin{equation}\label{eq:12}
\mathsf{1|45|3},\quad \mathsf{1|45|46},\quad \mathsf{25|45|3},\quad \mathsf{25|45|46}.
\end{equation}
The number of tag particles that can be added (at the same position in the same host) is only limited by the size of the system.
For example, two tags inside the first host reads $\mathsf{1|4545|3}$.
The minimum number of tiles \textsf{v} or \textsf{w} between two hosts can be two as in \textsf{13vw146}, one as in \textsf{13v13}, or zero as in \textsf{146253}.

The activation of every host particle extends the volume between the piston by the amount $2V_\mathrm{t}$, where
\begin{equation}\label{eq:3} 
V_\mathrm{t}\doteq V_\mathrm{f}-V_\mathrm{c}
=\sqrt{\sigma_\mathrm{L}\sigma_\mathrm{S}}
-\sqrt{H(\sigma_\mathrm{L}+\sigma_\mathrm{S}-H)}.
\end{equation}
Placing a tag, by contrast, does not change the volume.

The activation energy $\epsilon_m$ assigned to a particle from species $m$ pertains to the state of random agitation before jamming.
It consists of work against the ambient pressure exerted by the pistons and gravitational potential energy, all relative to the pseudo-vacuum. 
We are free to choose the mass density of small and large disks. Therefore, the gravitational potential energies $\gamma_\mathrm{L}$ and $\gamma_\mathrm{S}$ are independent parameters as is the expansion work $2pV_\mathrm{t}$.

A fourth parameter is the intensity of random agitations, denoted $T_\mathrm{k}$ in analogy to the intensity $k_\mathrm{B}T$ of thermal fluctuations.
All results coming out of configurational statistics will only depend on the following three (dimensionless) ratios of these four energetic parameters:
\begin{equation}\label{eq:4} 
\beta\doteq\frac{2pV_\mathrm{t}}{T_\mathrm{k}},\quad 
\Gamma_\mathrm{L}\doteq\frac{\gamma_\mathrm{L}}{2pV_\mathrm{t}},\quad
\Gamma_\mathrm{S}\doteq\frac{\gamma_\mathrm{S}}{2pV_\mathrm{t}}.
\end{equation}
The particle composition of a macrostate thus depends on the energetics prior to jamming.
The multiplicity of microstates that characterize a macrostate of given particle composition is the solution of a combinatorial problem.

%
\section{Combinatorics}\label{sec:comb}
%
The quasi-particles identified in Table~\ref{tab:t3} are statistically interacting in the sense that activating one particle affects the number $d_m$ of remaining slots for the activation of further particles from each species.
This effect can be accounted for by a generalized Pauli principle \cite{Hald91a} in the form \cite{copic,janac1},
\begin{equation}\label{eq:5}
  d_m =A_m-\sum_{m'=1}^M g_{mm'}(N_{m'}-\delta_{mm'}),
\end{equation}
with capacity constants $A_m$ and statistical interaction coefficients $g_{mm'}$ as listed in Table~\ref{tab:t4}, and where $N_m$ is the number of activated particles from species $m$.

\begin{table}[b]
  \caption{Capacity constants $A_m$ and statistical interaction coefficients $g_{mm'}$ for the particles from Table~\ref{tab:t3}.}\label{tab:t4}
\begin{center}
\begin{tabular}{c|c} \hline\hline
$m$~~ & ~~$A_m$  \\ \hline
1~~ & ~~$N-2$ \\
2~~ & ~~$N-3$ \\
3~~ & ~~$N-2$ \\
4~~ & ~~$N-3$ \\
5~~ & ~~$0$ \\ \hline\hline 
\end{tabular} \hspace{5mm}
\begin{tabular}{c|rrrrr} \hline\hline 
$g_{mm'}$ & $1$ & $2$ & $3$&4&5 \\ \hline 
$1$ & 2 & 2 & 1 & 2 &1\\ 
$2$ & 1 & 2 & 1& 1 &1\\ 
$3$ & 2 & 2& 2&2&1\\
$4$ & 1 & 2 & 1&2&1\\
$5$ & ~$-1$ & $-1$ & $-1$&$-1$&~~0\\ \hline\hline
\end{tabular}
\end{center}
\end{table}

The initial capacity for hosts grows linearly with the number of disks in the channel.
It is zero for tags, which can only be activated inside hosts.
Activating a host $(m'=1,\ldots,4)$ removes one or two slots for activating a further host $(m=1,\ldots,4)$ but adds one slot for activating a tag $(m=5)$. 
Activating a tag $(m'=5)$ removes one slot for activating hosts $(m=1,\ldots,4)$ but leaves the number of slots for activating a further tag $(m=5)$ invariant.

The multiplicity of jammed microstates with particle content $\{N_m\}$ is a product of binomials:
\begin{equation}\label{eq:6} 
W(\{N_m\})=\prod_{m=1}^M\left(\begin{array}{c}d_m+N_m-1 \\ N_m\end{array}\right). 
\end{equation}
We now have all the ingredients for the statistical mechanical analysis as was developed previously \cite{Wu94,Isak94,janac1}.

%
\section{Statistical mechanics}\label{sec:stat}
%
We can express both the excess volume and the entropy as functions of the average particle content $\{\langle N_m\rangle\}$ of a jammed macrostate as follows \cite{Isak94}:
\begin{equation}\label{eq:7} 
V-V_\mathrm{pv}=\sum_{m=1}^M\langle N_m\rangle \Delta V_m,
\end{equation}
\begin{subequations}\label{eq:8}
\begin{align}\label{eq:8a}
S &= k_B\sum_{m=1}^M\Big[\big(\langle N_{m}\rangle
+Y_m\big)\ln\big(\langle N_m\rangle+Y_m\big) \nonumber \\
 &\hspace{24mm} -\langle N_m\rangle \ln \langle N_m\rangle -Y_m\ln Y_m\Big],\\ \label{eq:8b}
Y_m &\doteq A_m-\sum_{m'=1}^Mg_{mm'} \langle N_{m'}\rangle.
\end{align}
\end{subequations}
The average particle numbers are the solutions of the linear equations \cite{Wu94,Isak94},
\begin{equation}\label{eq:9}
w_m \langle N_m\rangle+\sum_{m'=1}^Mg_{mm'} \langle N_{m'}\rangle =A_m.
 \end{equation} 
The auxiliary quantities $w_m$ are the non-negative set of solutions of the algebraic equations \cite{Wu94,Isak94},
\begin{equation}\label{eq:10} 
e^{\epsilon_m/T_\mathrm{k}}=(1+w_m)\prod_{m'=1}^M \big(1+w_{m'}^{-1}\big)^{-g_{m'm}}.
\end{equation}
The only parts that depend on the jamming protocol, i.e. on the three dimensionless parameters (\ref{eq:4}) are the exponents on the left.
The analytic solution of Eqs.~(\ref{eq:10}) is, in general, too unwieldy for display.
Its availability gives us explicit expressions for the scaled excess volume, the scaled entropy, and the particle densities,
\begin{equation}\label{eq:13} 
\bar{V}\doteq \frac{V-V_\mathrm{pv}}{2NV_\mathrm{t}},\quad
\bar{S}\doteq \frac{S}{Nk_\mathrm{B}},\quad
\bar{N}_m\doteq\frac{\langle N_m\rangle}{N},
\end{equation}
as functions of $\beta, \Gamma_\mathrm{L}, \Gamma_\mathrm{S}$, representing, as noted in the context of (\ref{eq:4}), the (inverse) intensity of random agitations and the gravitational potential energies of large and small disks, all in units of expansion work against the pistons before jamming.

%
\section{Weights in balance}\label{sec:we-in-ba}
%
We begin our presentation of results at the border between the two regimes of heavy and light large disks.
When we declare
\begin{equation}\label{eq:14}
\Gamma_\mathrm{L}=\Gamma_\mathrm{S}\doteq\Gamma
 \end{equation} 
to hold, it means that the gravitational potential energies of large and small disks are equal when they touch the same wall.
This equality removes the effect of gravity from the activation energies of the particles from species $m=2, 3, 5$.
On either side of this border, gravity favors qualitatively different jamming patterns as we shall see.
Neither matches the pattern right on the border.

\begin{figure}[htb]
  \begin{center}
\includegraphics[width=43mm]{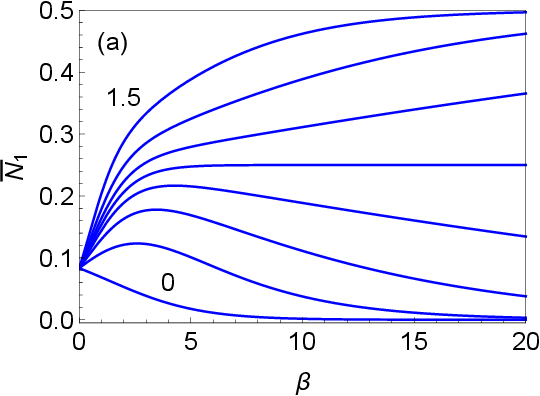}%
\hspace*{1mm}\includegraphics[width=43mm]{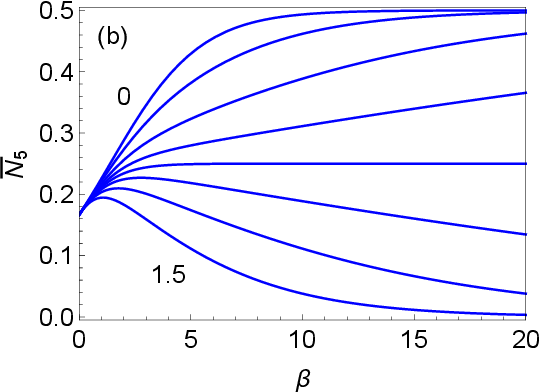}
\includegraphics[width=43mm]{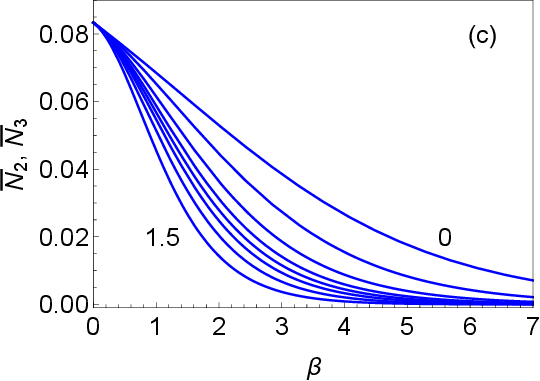}%
\hspace*{1mm}\includegraphics[width=43mm]{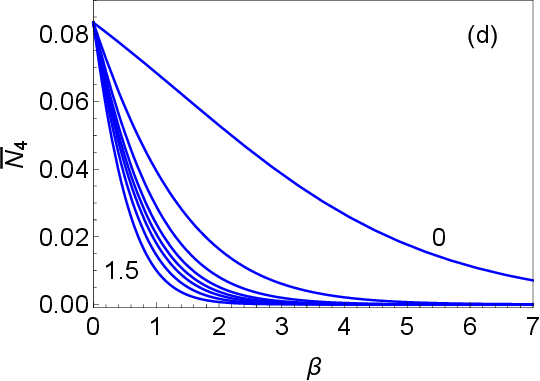}
\end{center}
\caption{Population densities (a) $\bar{N}_1$, (b) $\bar{N}_5$, (c) $\bar{N}_2=\bar{N}_3$, (d) $\bar{N}_4$ versus parameter $\beta$ for fixed parameter values $\Gamma=0$, 0.5, 0.75, 0.9, 1.0, 1.1, 1.25, 1.5.}
  \label{fig:f1}
\end{figure}

In Fig.~\ref{fig:f1} we show the population densities $\bar{N}_m$ plotted versus $\beta$ for all five species of particles at different values of $\Gamma$. 
Recall that increasing $\beta$ means reducing the intensity of random agitations before jamming and increasing $\Gamma$ means increasing the effects of gravity (by reducing the ambient pressure before jamming)  \cite{note1}.

In the high-intensity limit, we have
\begin{equation}\label{eq:18}
\bar{N}_1=\cdots=\bar{N}_4=\frac{1}{12},\quad \bar{N}_5=\frac{1}{6}\quad  :~ \beta=0.
\end{equation}
The contributions of species $m=2,3,4$ quickly fade away as $\beta$ increases from zero. 
That trend is aided by gravity.
The two dominant species, $m=1,5$, are a host/tag pair with tags 5 existing inside hosts 1 in arbitrary numbers and with only hosts contributing to excess volume.

For weak gravity, $\Gamma<1$, the host population, $\bar{N}_1$, initially increases and then approaches zero as $\beta$ increases.
For strong gravity, $\Gamma>1$, it increases monotonically toward the asymptotic value, $\bar{N}_1=\frac{1}{2}$. 
The trends are roughly opposite for the tag population density, $\bar{N}_5$.
At the border between these two regimes, $\Gamma=1$, both population densities level off at $\bar{N}_1=\bar{N}_5=\frac{1}{4}$ in the low-intensity limit $\beta\to\infty$.

\begin{figure}[htb]
  \begin{center}
\includegraphics[width=43mm]{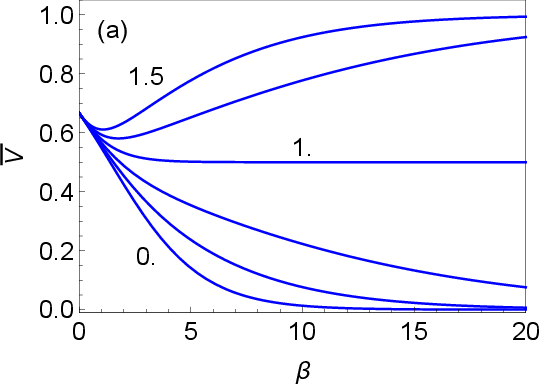}%
\hspace*{1mm}\includegraphics[width=43mm]{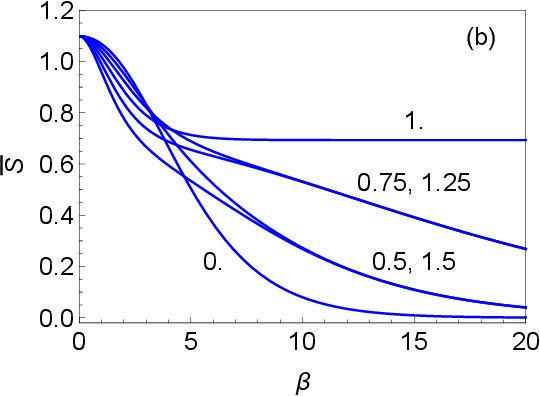}
\end{center}
\caption{ (a) Excess volume $\bar{V}$ and (b) entropy $\bar{S}$ versus parameter $\beta$ for fixed values $\Gamma=0$, 0.5, 0.75, 1.0, 1.25, 1.5.}
  \label{fig:f2}
\end{figure}

The effects of particle population densities on the excess volume and entropy of jammed macrostates are governed by Eqs.~(\ref{eq:7}) and (\ref{eq:8}), respectively, and shown in Fig.~\ref{fig:f2} (for fewer values of $\Gamma$).
In the high-intensity limit, we thus infer from (\ref{eq:18}) the values, 
\begin{equation}\label{eq:27} 
\bar{V}=\frac{2}{3},\quad \bar{S}=\ln3\quad :~ \beta=0,
\end{equation}
for excess volume and entropy of the most disordered macrostate.

As $\beta$ increases, the excess volume initially goes down no matter whether gravity is weak or strong [panel (a)].
If gravity is weak it continues to go down and vanishes aymptotically for $\beta\to\infty$.
For strong gravity, on the other hand, the downward trend is quickly reversed into an upward trend toward the asymptotic value $\bar{V}=1$.

Irrespective of whether the intensity reduction of random agitations before jamming leads to larger or smaller volume, it always leads to lower entropy [panel(b)].
The entropy approaches zero at rates which vary a great deal with $\Gamma$. 
The ordered state being approached for $\Gamma<1$ is different from the one approached for $\Gamma>1$.
The quickest approach by far to the ordered jammed state pertains to the zero-gravity case. 

The borderline case, $\Gamma=1$, is exceptional. 
The intensity reduction does not lead to an ordered jammed state.
The excess volume settles at an intermediate value and the entropy at a nonzero value:
\begin{equation}\label{eq:28}
\lim_{\beta\to\infty}\bar{V}=\frac{1}{2},\quad 
\lim_{\beta\to\infty}\bar{S}=\ln2\quad :~ \Gamma=1.
\end{equation}

\begin{figure}[t]
  \begin{center}
\includegraphics[width=60mm]{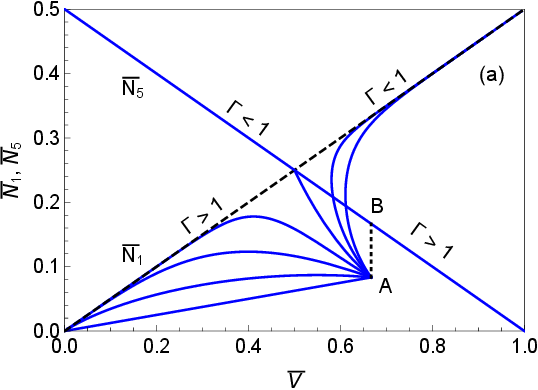}
\includegraphics[width=43mm]{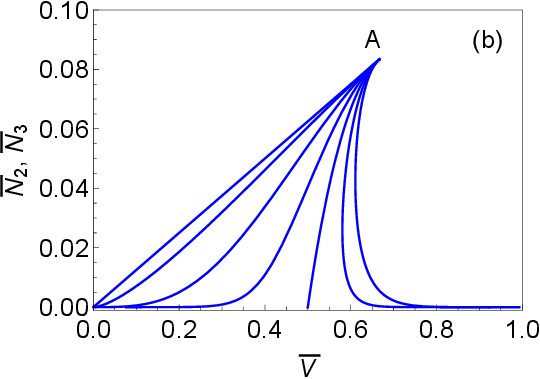}%
\hspace*{1mm}\includegraphics[width=43mm]{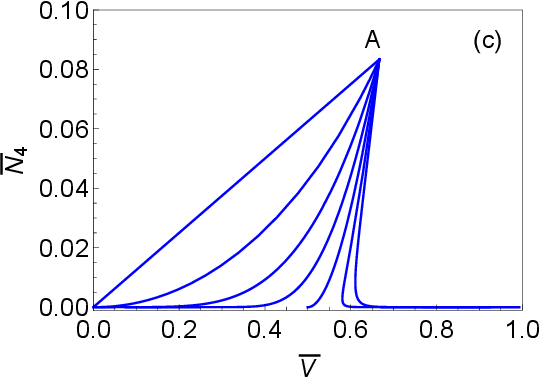}
\end{center}
\caption{Population densities $\bar{N}_m$, $m=1,\ldots,5$ versus excess volume $\bar{V}$ for fixed values $\Gamma=0$, 0.25, 0.5, 0.75, 1.0, 1.25, 1.5 from bottom up in panel (a) and from left to right in panels (b) and (c). The dotted line connects the points pertaining to $\beta=0$ for hosts and tags. The dashed diagonal represents the sum $\bar{N}_1+\bar{N}_2+\bar{N}_3+\bar{N}_4$.}
  \label{fig:f3}
\end{figure}

In order to better understand the roles of the different particle species, we plot the population densities versus volume [Fig.~\ref{fig:f3}].
In these parametric plots, the high-intensity limit $\beta=0$ is realized at point $A$ for hosts and point $B$ for tags.
Increasing $\beta$ toward infinity means moving along any path toward one or the other of two corners in each panel, depending on whether $\Gamma\gtrless1$. 
In the exceptional case $\Gamma=1$, all paths stop at $\bar{V}=\frac{1}{2}$.

An important fact about the case (\ref{eq:14}) of balanced gravity is that the overall population density of particles is an invariant:
\begin{equation}\label{eq:19}
\bar{N}_\mathrm{tot}\doteq\sum_{m=1}^5\bar{N}_m=\frac{1}{2}=\mathrm{const}.
\end{equation}
This is illustrated by the two diagonal lines in Fig.~\ref{fig:f3}(a), where the dashed line represents the population density of all hosts combined.
Recall that the tags $m=5$ do not contribute to excess volume.
Their numbers are determined, nevertheless, by the excess volume via the conservation law.
Any increase in volume caused by the activation of one or the other host necessarily crowds out one tag.
The curves in Fig.~\ref{fig:f3} again highlight the observation made earlier that hosts $m=2,3,4$ only contribute significantly at high intensity.

It is instructive to take a closer look at the three distinct macrostates associated with $\beta=\infty$.
All have $\bar{N}_2=\bar{N}_3=\bar{N}_4=0$.

(i) $\Gamma<1$: The state with $\bar{N}_1=0$ and $\bar{N}_5=\frac{1}{2}$ is realized and has $\bar{V}=0$, $\bar{S}=0$.
It is a doublet consisting of the reference state (\ref{eq:2}) and the state $\mathsf{14545\cdots4546}$.
The former is free of particles.  The latter contains one host 2 with a saturated number of tags 5 inside. 

(ii) $\Gamma>1$: The state with $\bar{N}_1=\frac{1}{2}$ and $\bar{N}_5=0$ is realized and has $\bar{V}=1$, $\bar{S}=0$.
It is a singlet packed with hosts 1: $\mathsf{13vw13vw\cdots13v}$. Hosts 1 proliferate because they have negative activation energies.

(iii) $\Gamma=1$: The state with $\bar{N}_1=\bar{N}_5=\frac{1}{4}$ is realized and has $\bar{V}=\frac{1}{2}$, $\bar{S}=\ln2$.
It is highly degenerate. The hosts 1 are randomly distributed between vacuum tiles with a random number of tags inside. Hosts 1 and tags 5 have zero activation energies, whereas the hosts, 2, 3, 4 have positive activation energies.

\begin{figure}[htb]
  \begin{center}
\includegraphics[width=60mm]{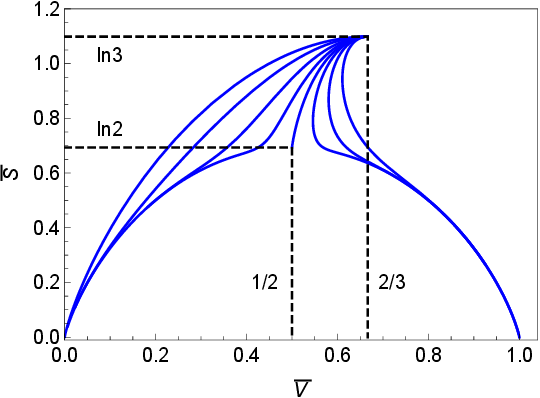}
\end{center}
\caption{ Entropy $\bar{S}$ versus excess volume $\bar{V}$ for fixed values $\Gamma=0$, 0.5, 0.75, 0.9, 1.0, 1.1, 1.25, 1.5 (from left to right).}
  \label{fig:f4}
\end{figure}

With this information we are ready to interpret a graphical representation of entropy versus excess energy [Fig.~\ref{fig:f4}].
The parameter $\beta$ runs from zero to infinity along each path from the top down.
All paths start at coordinates (\ref{eq:27}).
Paths for $\Gamma<1$ end at $\bar{V}=0$, $\bar{S}=0$, and paths for $\Gamma>1$ at $\bar{V}=1$, $\bar{S}=0$.
At critical gravity, $\Gamma=1$, both volume, the path ends at coordinates (\ref{eq:28}).

Compact analytic expressions for the curves in Figs.~\ref{fig:f3} and \ref{fig:f4} are available for the cases of zero and critical gravity.
In the case $\Gamma=0$ we have
\begin{subequations}\label{eq:20}
\begin{align}\label{eq:20a}
&\bar{N}_1=\cdots=\bar{N}_4=\frac{1}{8}\bar{V},\quad 
\bar{N}_5=\frac{1}{2}\big(1-\bar{V}\big), \\ \label{eq:20b}
&\bar{S}=\big(\bar{V}-1\big) \ln \left(\frac{2}{\bar{V}}-2\right)+\ln
   \left(\frac{2}{\bar{V}}\right),
\end{align}
\end{subequations}
where the range of volume is $0\leq\bar{V}\leq\frac{2}{3}$.
In the case $\Gamma=1$ we have
\begin{subequations}\label{eq:21}
\begin{align}\label{eq:21a}
& \bar{N}_1=\frac{(\bar{V}-1)^2}{2\bar{V}},\quad 
\bar{N}_2=\bar{N}_3=\frac{3}{2}-\bar{V}-\frac{1}{2\bar{V}}, \\ \label{eq:21b}
& \bar{N}_4=2\bar{V}+\frac{1}{2\bar{V}}-2,\quad 
\bar{N}_5=\frac{1}{2}\big(1-\bar{V}\big), \\ \label{eq:21c}
& \bar{S}=2 (\bar{V}-1) \ln \left(\frac{1-\bar{V}}{2
   \bar{V}-1}\right)-\ln (2 \bar{V}-1),
\end{align}
\end{subequations}
across the more restricted range $\frac{1}{2}\leq\bar{V}\leq\frac{2}{3}$ of volume.
A brief account of the weights-in-balance case $\Gamma_\mathrm{L}=\Gamma_\mathrm{S}$ including versions of Figs.~\ref{fig:f3} and \ref{fig:f4} was reported in \cite{LM19}.

%
\section{Heavy large disks}\label{sec:he-la-di}
%
Next we discuss the case where the large disks are heavier than the small disks in the sense that
\begin{equation}\label{eq:15}
\Gamma_\mathrm{L}>\Gamma_\mathrm{S}
 \end{equation} 
All graphs for this case use $\Gamma_\mathrm{L}=2\Gamma_\mathrm{S}$.
In the following, we merely point out the main differences from the weights-in-balance case.

\begin{figure}[b]
  \begin{center}
\includegraphics[width=43mm]{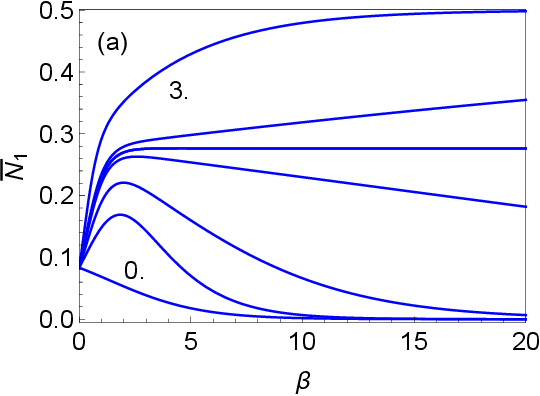}%
\hspace*{1mm}\includegraphics[width=40mm]{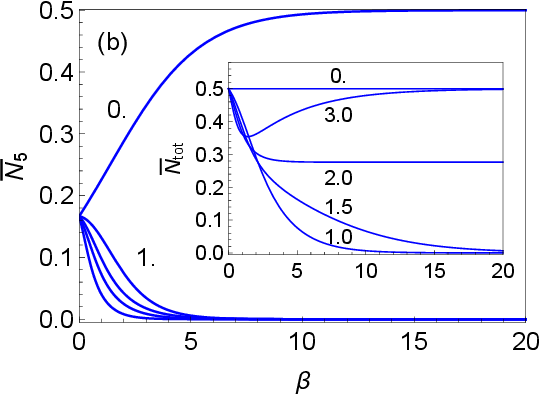}
\includegraphics[width=43mm]{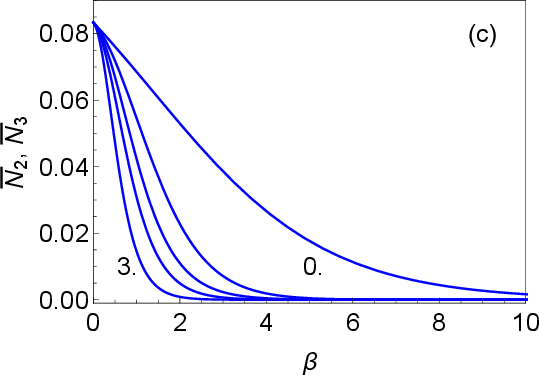}%
\hspace*{1mm}\includegraphics[width=43mm]{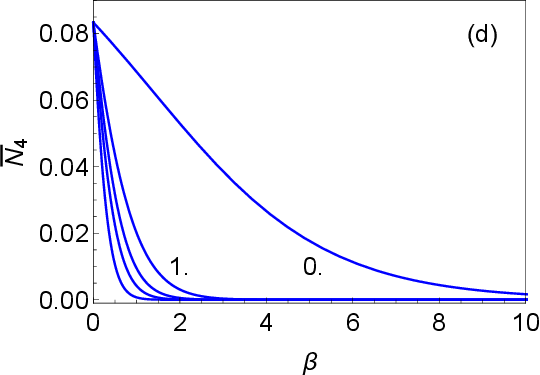}
\end{center}
\caption{Population densities (a) $\bar{N}_1$, (b) $\bar{N}_5$, (c) $\bar{N}_2=\bar{N}_3$, (d) $\bar{N}_4$ versus parameter $\beta$ for fixed values $\Gamma_\mathrm{L}=0$, 1.0, 1.5, [1.9], 2.0, [2.1], 3.0. The inset shows $\bar{N}_\mathrm{tot}$ versus $\beta$. }
  \label{fig:f5}
\end{figure}

The particle population densities $\bar{N}_m$ plotted versus $\beta$ [Fig.~\ref{fig:f5}] look similar with one notable exception. 
The role of tags is greatly diminished for any $\Gamma_\mathrm{L}>0$.
This has the consequence that jamming with reduced intensity of random agitations produces macrostates dominated by hosts 1 alone.
There are again regimes of strong and weak gravity with criticality now at $\Gamma_\mathrm{L}=2$.
We find 
\begin{equation}\label{eq:22} \rule[-2mm]{0mm}{6mm}
\lim_{\beta\to\infty} \bar{N}_1=\left\{
\begin{array}{cl} 0 &:~ \Gamma_\mathrm{L}<2, \\ \rule[-2mm]{0mm}{6mm}
\frac{1}{10}(5-\sqrt{5}) &:~ \Gamma_\mathrm{L}=2, \\ \rule[-2mm]{0mm}{6mm}
\frac{1}{2} &:~ \Gamma_\mathrm{L}>2. \\
\end{array} \right.
\end{equation}
The overall particle content was the invariant (\ref{eq:19}) for weights in balance.
This quantity now varies with $\beta$ as shown in the inset to Fig.~\ref{fig:f5}(b).
Only for $\Gamma_\mathrm{L}=0$ is it still an invariant.
Turning on gravity precipitates significant changes.

For weak gravity, $0<\Gamma_\mathrm{L}<2$, reducing the intensity of random agitations has the consequence of depleting the system of particles gradually and completely.
Asymptotically for $\beta\to\infty$, the jammed macrostate becomes the pseudo-vacuum.
At strong gravity, $\Gamma_\mathrm{L}>2$, the particle content also begins to go down but then turns around and returns to the initial value asymptotically for $\beta\to\infty$.
Initially, all particles are present in the proportions (\ref{eq:18}), finally it's all hosts 1.
For the case $\Gamma_\mathrm{L}=2$, the particle content goes down as well, but levels off as stated in (\ref{eq:22}), again representing hosts 1 alone.

\begin{figure}[t]
  \begin{center}
\includegraphics[width=43mm]{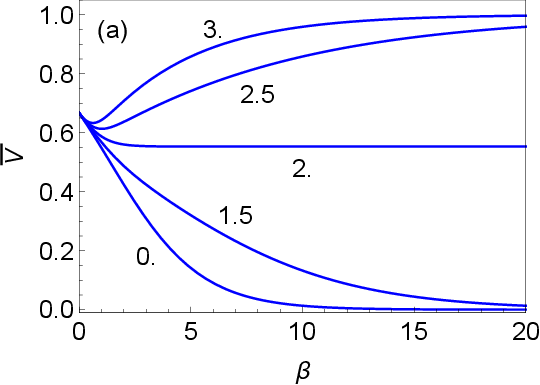}%
\hspace*{1mm}\includegraphics[width=43mm]{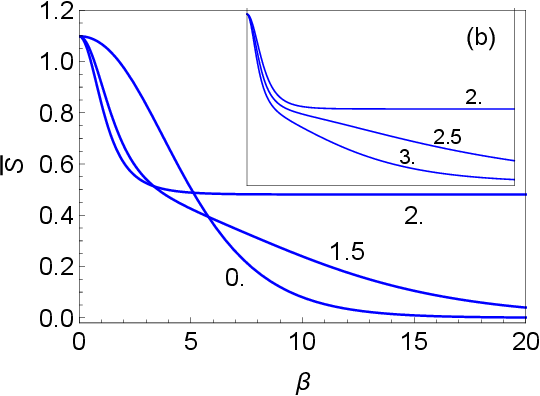}
\end{center}
\caption{ (a) Excess volume $\bar{V}$ and (b) entropy $\bar{S}$ versus parameter $\beta$ for fixed values $\Gamma_\mathrm{L}$ as indicated.}
  \label{fig:f6}
\end{figure}

The dependence on $\beta$ of volume and entropy remain qualitatively the same [Fig.~\ref{fig:f6}] as before, but with different asymptotics at the border $\Gamma_\mathrm{L}=2$ between the two regimes:
\begin{subequations}\label{eq:23-24}
\begin{equation}\label{eq:23}
\lim_{\beta\to\infty} \bar{V}=1-\frac{1}{\sqrt{5}}\simeq 0.553
\quad :~ \Gamma_\mathrm{L}=2,
\end{equation}
\begin{equation}\label{eq:24}
\lim_{\beta\to\infty} \bar{S}=\frac{1}{2} 
\ln \left(\frac{1}{2} \left(3+\sqrt{5}\,\right)\right)\simeq 0.481.
\end{equation}
\end{subequations}

\begin{figure}[b]
  \begin{center}
\includegraphics[width=43mm]{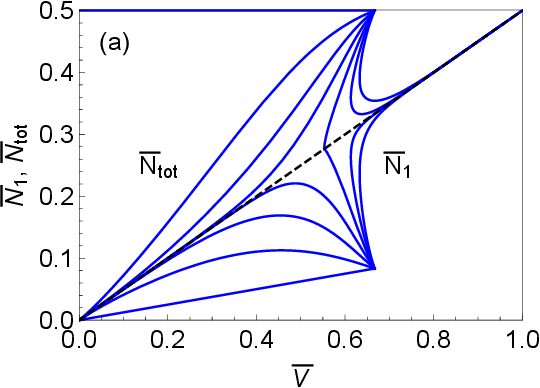}%
\hspace*{1mm}\includegraphics[width=43mm]{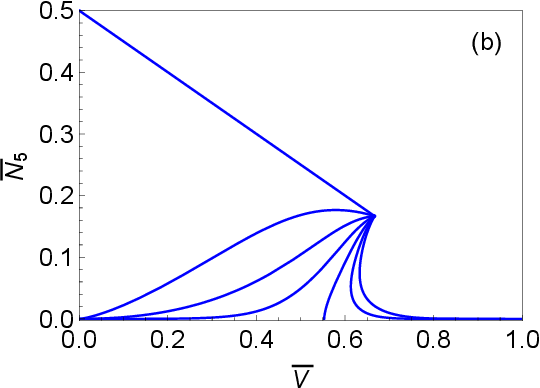}
\includegraphics[width=43mm]{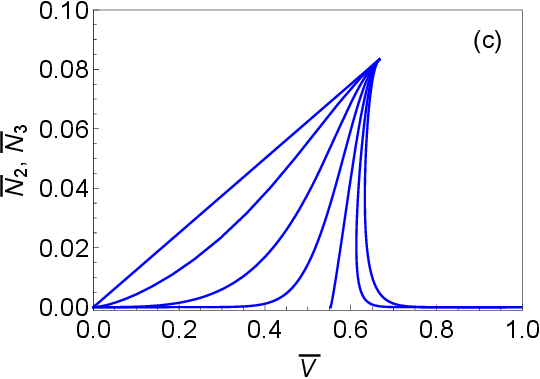}%
\hspace*{1mm}\includegraphics[width=43mm]{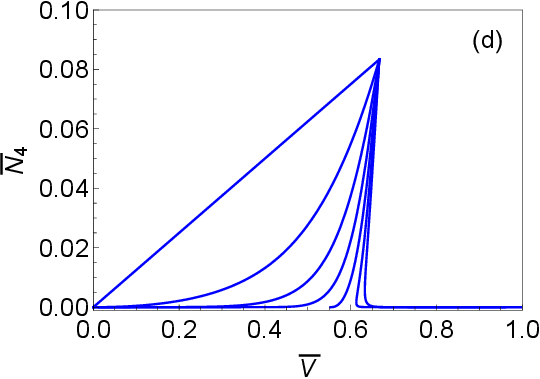}
\end{center}
\caption{Population densities $\bar{N}_m$, $m=1,\ldots,5$ and $\bar{N}_\mathrm{tot}$ versus excess volume $\bar{V}$ for fixed values $\Gamma_\mathrm{L}=0$, 0.5 1.0, 1.5, 2.0, 2.5, 3.0 [$\bar{N}_1$ from bottom up, $\bar{N}_\mathrm{tot}, \bar{N}_5$ from top down, $\bar{N}_2, \bar{N}_3, \bar{N}_4$ from left to right.] The dashed diagonal represents the sum $\bar{N}_1+\bar{N}_2+\bar{N}_3+\bar{N}_4$.}
  \label{fig:f7}
\end{figure}

The diminished role of tags 5 is again highlighted in Fig.~\ref{fig:f7}, where the population densities are plotted versus excess volume $\bar{V}$.
Panel (a) also shows the overall particle content.
Along two of the curves there are short segments along which the system loses particles while the volume of the jammed state increases.
Changes occur also in two of the macrostates associated with $\beta=\infty$ (all with $\bar{N}_2=\bar{N}_3=\bar{N}_4=\bar{N}_5=0$).

(i) $0<\Gamma_\mathrm{L}<2$:
The state realized is the reference state, which (by construction) contains no particles and has $\bar{V}=0$ and $\bar{S}=0$.

(ii) $\Gamma_\mathrm{L}>2$:
The state realized is a close-packed array of hosts, implying $\bar{N}_1=\frac{1}{2}$, $\bar{V}=1$, and $\bar{S}=0$ as we have encountered in the strong-gravity regime (Sec.~\ref{sec:we-in-ba}).

(iii) $\Gamma_\mathrm{L}=2$:
The state realized is again highly degenerate and thus carries entropy but unlike in Sec.~\ref{sec:we-in-ba} it contains only hosts.
Their average density is higher, $\bar{N}_1\simeq0.276$, but they are shorter for lack of tags.

\begin{figure}[htb]
  \begin{center}
\includegraphics[width=60mm]{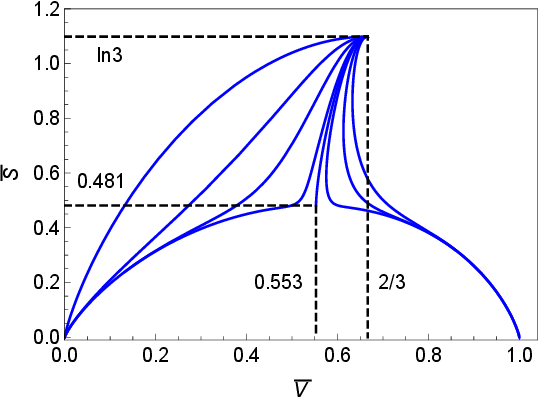}
\end{center}
\caption{ Entropy $\bar{S}$ versus excess volume $\bar{V}$ for fixed values $\Gamma_\mathrm{L}=0$, 1.0, 1.5, 1.8, 2.0, 2.2, 2.5, 3.0 (from left to right).}
  \label{fig:f8}
\end{figure}

It is perhaps surprising that all the changes we have noted between this case of heavy large disks and the previous case of weights in balance produce a mere distortion when we plot entropy versus volume [Fig.~\ref{fig:f8}].
One of the landmarks in the plot has the same coordinates (\ref{eq:27}), the other is somewhat shifted from (\ref{eq:28}) to (\ref{eq:23-24}).

%
\section{Light large disks}\label{sec:li-la-di}
%
If the weights of large and small disks are oppositely unbalanced, in the sense that
\begin{equation}\label{eq:16}
\Gamma_\mathrm{L}<\Gamma_\mathrm{S},
 \end{equation} 
then jamming produces yet different results.
There are two regimes with different ordering patterns.
We first discuss the borderline case, which involves subtle limits, and then one case on either side of the border.

\subsection{$\Gamma_\mathrm{L}=\frac{1}{2}\Gamma_\mathrm{S}$.}\label{sec:1-over-2}
\begin{figure}[htb]
  \begin{center}
\includegraphics[width=43mm]{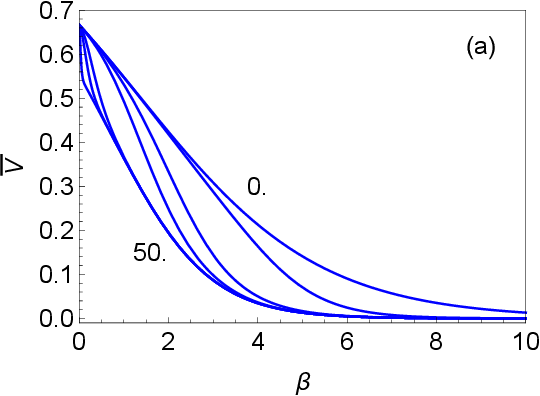}%
\hspace*{1mm}\includegraphics[width=43mm]{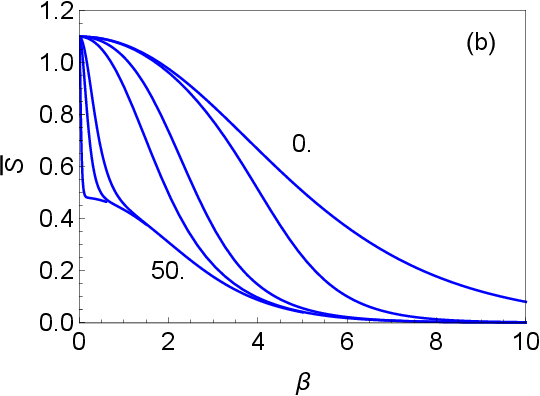}
\end{center}
\caption{ (a) Excess volume $\bar{V}$ and (b) entropy $\bar{S}$ versus $\beta$ for fixed values $\Gamma_\mathrm{L}=0$, 0.1, 0.5, 1, 5, 10, 50.}
  \label{fig:f10}
\end{figure}

The hallmark of this case is the presence of distinct weak-gravity and strong-gravity jamming patterns with a crossover between them upon variation of the intensity of random agitations before jamming. 
The clearest evidence is seen in plots of $\bar{V}$ and $\bar{S}$ versus $\beta$ [Fig.~\ref{fig:f10}].

In both panels the top curve represents zero gravity and the bottom curve very strong gravity.
For intermediate strengths of gravity, the curves initially follow the zero-gravity behavior and then, with increasing $\beta$, cross over to the strong gravity behavior.
For large values of $\Gamma_\mathrm{L}$, the crossover turns into a transition that destabilizes the zero-gravity behavior for all $\beta>0$. 
A discontinuity emerges in the combined but non-interchangeable limits $\Gamma_\mathrm{L}\to\infty$, $\beta\to0$ between the most disordered macrostate with values (\ref{eq:27}),
\begin{subequations}\label{eq:25}
\begin{equation}\label{eq:25a}
\lim_{\Gamma_\mathrm{L}\to\infty}\lim_{\beta\to0} \bar{V}=\frac{2}{3}\simeq 0.667,
\end{equation}
\begin{equation}\label{eq:25b}
\lim_{\Gamma_\mathrm{L}\to\infty}\lim_{\beta\to0} \bar{S}=\ln 3
\simeq 1.099,
\end{equation}
\end{subequations}
and the partially ordered macrostate with values,
\begin{subequations}\label{eq:26}
\begin{equation}\label{eq:26a}
\lim_{\beta\to0} \lim_{\Gamma_\mathrm{L}\to\infty}
\bar{V}=1-\frac{1}{\sqrt{5}}\simeq 0.553,
\end{equation}
\begin{equation}\label{eq:26b}
\lim_{\beta\to0} \lim_{\Gamma_\mathrm{L}\to\infty}
\bar{S}=\frac{1}{2} 
\ln \left(\frac{1}{2} \left(3+\sqrt{5}\,\right)\right)\simeq 0.481.
\end{equation}
\end{subequations}

\begin{figure}[t]
  \begin{center}
\includegraphics[width=43mm]{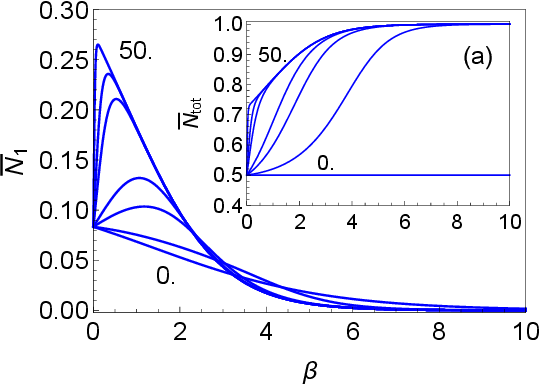}%
\hspace*{1mm}\includegraphics[width=43mm]{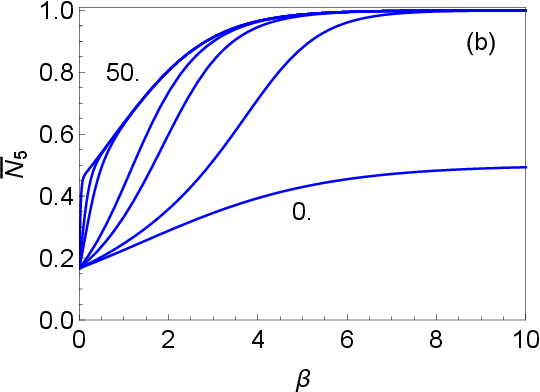}
\includegraphics[width=43mm]{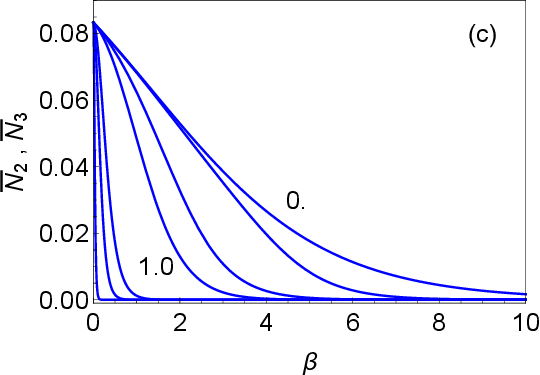}%
\hspace*{1mm}\includegraphics[width=43mm]{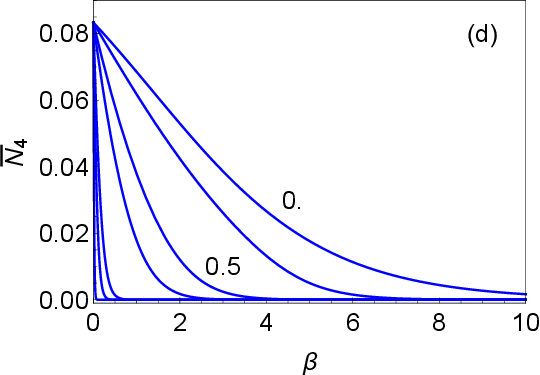}
\end{center}
\caption{Population densities (a) $\bar{N}_1$, (b) $\bar{N}_5$, (c) $\bar{N}_2=\bar{N}_3$, (d) $\bar{N}_4$ versus $\beta$ for fixed values $\Gamma_\mathrm{L}=0$, 0.1, 0.5, 1, 5, 10, 50.}
  \label{fig:f9}
\end{figure}

This two-stage process of ordering in the jammed macrostate produced by random agitations of diminishing intensity has its characteristic signatures in the particle population densities [Fig.~\ref{fig:f9}]. 
During the first stage, the populations of hosts 1 and tags 5 are enhanced, whereas hosts 2,3,4 become extinct.
During the second stage, hosts 1 disappear gradually whereas the population of tags 5 gradually increases toward its maximum value.
The particle population overall, $\bar{N}_\mathrm{tot}$, increases monotonically with increasing $\beta$ [inset]. 
With increasing $\Gamma_\mathrm{L}$, the first stage becomes more and more abrupt, whereas the second stage remains gradual.

\begin{figure}[htb]
  \begin{center}
\includegraphics[width=60mm]{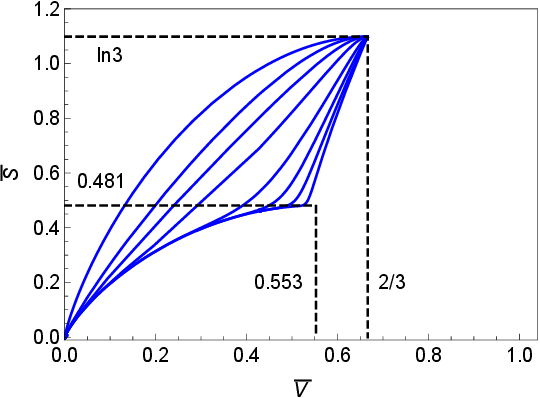}
\end{center}
\caption{ Entropy $\bar{S}$ versus excess volume $\bar{V}$ for fixed values $\Gamma_\mathrm{L}=0$, 0.5, 1, 2, 5, 10, 20, 50 (from left to right).}
  \label{fig:f12}
\end{figure}

The signature plot, $\bar{S}$ versus $\bar{V}$, for this case is shown in Fig.~\ref{fig:f12}.
The dashed lines from Fig.~\ref{fig:f8}, remain relevant.
Weak gravity initially has roughly the same effect in the two cases represented by Figs.~\ref{fig:f8} and \ref{fig:f12}. 
For light large disks and infinitely strong gravity, ${\Gamma_\mathrm{L}\to\infty}$, the path of macrostates during the first stage quickly transitions from coordinates (\ref{eq:25}) to (\ref{eq:26}).
The latter are identical to coordinates (\ref{eq:23-24}) in Fig.~\ref{fig:f8}, where they specify a macrostate in the limit $\beta\to\infty$.

The macrostate with $\bar{V}=0$ and $\bar{S}=0$ arrived at in the limit $\beta\to\infty$ is a singlet made of one host 2 packed with tags 5: $\mathsf{14545\cdots4546}$.
The macrostate (\ref{eq:26}) contains hosts 1 with $\bar{N}_1\simeq0.276$ randomly distributed between vacuum tiles, and tags 5 with $\bar{N}_5\simeq0.447$ randomly distributed among the hosts. 

The two macrostates with the same coordinates, (\ref{eq:23-24}) in Fig.~\ref{fig:f8} and (\ref{eq:26}) in Fig.~\ref{fig:f12}, are very different in particle composition.
The former only contains hosts 1, whereas the latter contains hosts 1 and tags 5. The former (latter) is realized for very weak (strong) random agitations before jamming.

How do we explain that these two macrostates with different particle composition have the same excess volume and the same entropy?
Only hosts carry excess volume.
Both macrostates contain the same number of hosts, hence produce the same $\bar{V}$.
Hosts alone are smaller in size and the size is unique. 
Hosts with tags are larger in size but the sizes are many. 
Hosts with tags have fewer options for placement between them, which subtracts entropy, but they produce distinguishable permutations, which adds entropy.
This is no explanation, but an argument in support of plausibility.

\subsection{$\Gamma_\mathrm{L}=\frac{1}{3}\Gamma_\mathrm{S}$.}\label{sec:1-over-3}
\begin{figure}[htb]
  \begin{center}
\includegraphics[width=43mm]{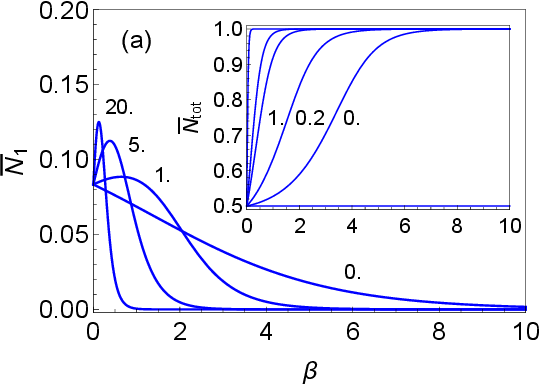}%
\hspace*{1mm}\includegraphics[width=43mm]{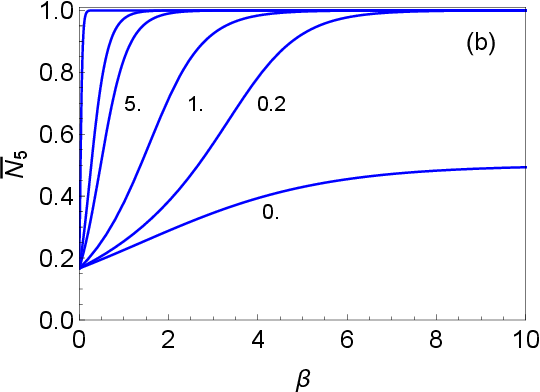}
\end{center}
\caption{Population densities versus $\beta$ (a) $\bar{N}_1$ for $\Gamma_\mathrm{S}=0$, 1, 5, 20, (b) $\bar{N}_5$ for $\Gamma_\mathrm{S}=0$, 0.2, 1, 5, 10, 100. The inset shows $\bar{N}_\mathrm{tot}$ for $\Gamma_\mathrm{S}=0$, 0.2, 1, 5, 10, 100.}
  \label{fig:f13}
\end{figure}

This case of very light large disks exhibits the least complex behavior.
It resembles the borderline case in some aspects and differs in others.
For $\Gamma_\mathrm{L}>0$, all host populations are suppressed eventually as $\beta\to\infty$ and the tag population grows to its maximum value [Fig.~\ref{fig:f13}].
Only hosts 1 show hints of a two-stage process emerging as $\Gamma_\mathrm{L}$ becomes large: a quick initial rise followed by an equally fast drop.
Compared to the case of Sec.~\ref{sec:1-over-2}, the second stage is now much faster in the event of strong gravity.
The variation of excess volume and entropy with $\beta$ (not shown) is equally fast and all downward. 

\begin{figure}[htb]
  \begin{center}
\includegraphics[width=60mm]{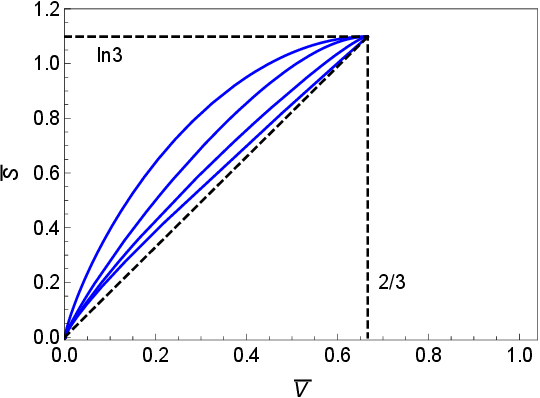}
\end{center}
\caption{ Entropy $\bar{S}$ versus excess volume $\bar{V}$ for $\Gamma_\mathrm{S}=0$, 1, 5, 100 (from left to right).}
  \label{fig:f16}
\end{figure}

The plot of $\bar{S}$ versus $\bar{V}$ [Fig.~\ref{fig:f16}] shows no complexity but is informative nevertheless.
We observe that the relation between excess volume and entropy becomes increasingly linear as gravity becomes stronger.
The only way to interpret this is that in the limit $\Gamma_\mathrm{L}\to\infty$ both quantities collapse discontinuously to zero as $\beta$ becomes nonzero.
The initial macrostate has maximum disorder. The final state is the same singlet as identified in Sec.~\ref{sec:1-over-2}.

\subsection{$\Gamma_\mathrm{L}=\frac{2}{3}\Gamma_\mathrm{S}$.}\label{sec:2-over-3}
On the other side of the borderline case, where the large disks are not much lighter than the small disks, the jamming patterns are more varied.
We can state (without showing evidence) that all particle population densities vary less precipitously with $\beta$ than in the preceding two cases.
However, turning on gravity produces an abrupt change in the (average) particle composition of the jammed macrostate. Further abrupt changes occur when gravity crosses criticality at $\Gamma_\mathrm{L}=2$.

\begin{figure}[htb]
  \begin{center}
\includegraphics[width=43mm]{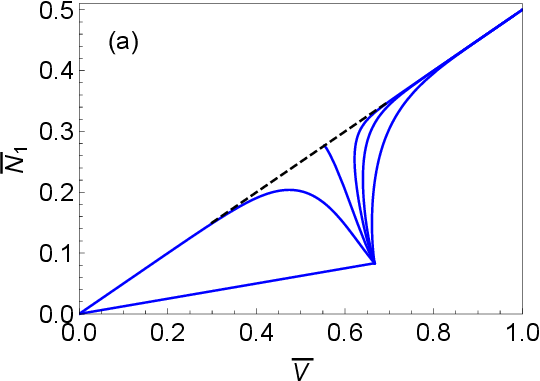}%
\hspace*{1mm}\includegraphics[width=43mm]{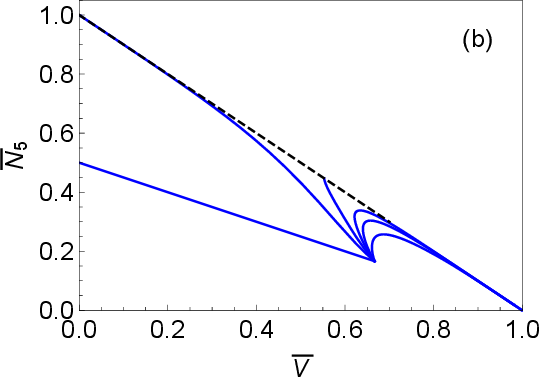}
\includegraphics[width=43mm]{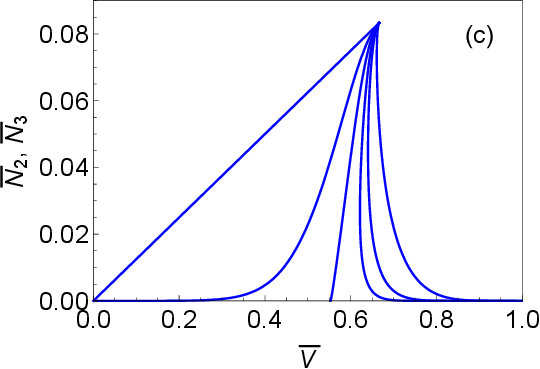}%
\hspace*{1mm}\includegraphics[width=43mm]{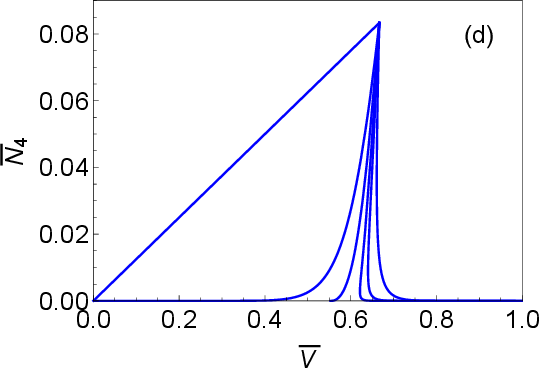}
\includegraphics[width=43mm]{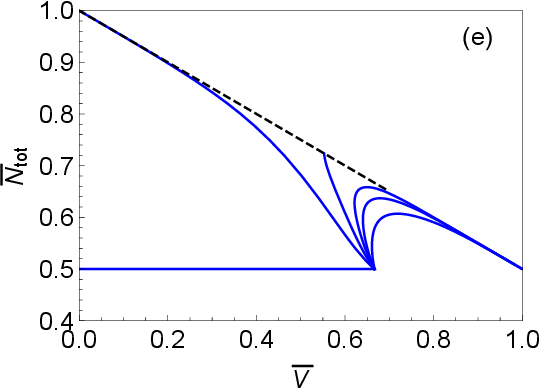}
\end{center}
\caption{Population densities $\bar{N}_m$, $m=1,\ldots,5$ and $\bar{N}_\mathrm{tot}$ versus excess volume $\bar{V}$ for fixed values $\Gamma_\mathrm{S}=0$, 2, 3, 4, 5, 10 (from left to right).}
  \label{fig:f19}
\end{figure}

These changes are best illustrated in plots of particle population densities versus excess volume [Fig.~\ref{fig:f19}]
We have noted before (Sec.~\ref{sec:we-in-ba}) that the jammed macrostate at zero gravity is a doublet in the limit $\beta\to\infty$. 
Increasing $\Gamma_\mathrm{L}$ from zero removes that degeneracy, which has the effect that the (average) tag population density doubles.
For $\Gamma_\mathrm{L}<2$ the effect of increasing $\beta$ is that the population of hosts 1 first increases and then decreases toward zero [panel (a)].
The populations of the other three hosts decrease from the start at $\beta=0$ [panels (c) and (d)] such that the excess volume is a monotonically decreasing function.  
The overall population density of particles, on the other hand, is monotonically increasing [panel (e)] on account of the population increase of tags 5 [panel (b)].

For critical gravity, $\Gamma_\mathrm{L}=2$, the effect of increasing $\beta$ is that the population densities of hosts 1 and tags 5 both increase whereas those of the remaining hosts vanish gradually as in all cases. 
The macrostate in the limit $\beta\to\infty$ is a familiar one, encountered before in the case of heavy large disks, namely the disordered state with the values (\ref{eq:23-24}) for excess volume and entropy.

\begin{figure}[htb]
  \begin{center}
\includegraphics[width=60mm]{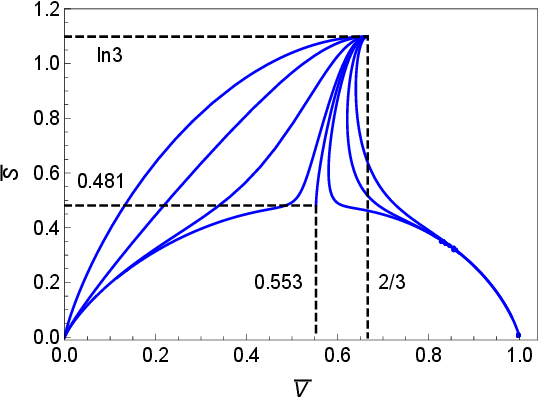}
\end{center}
\caption{ Entropy $\bar{S}$ versus excess volume $\bar{V}$ for fixed values $\Gamma_\mathrm{S}=0$, 1, 2, 2.8, 3, 3.2, 4, 5 (from left to right).}
  \label{fig:f20}
\end{figure}

For stronger gravity, $\Gamma_\mathrm{L}>2$, the particle composition of the jammed macrostate approaches an ordered state again in the limit $\beta\to\infty$.
This state is packed with hosts 1 and has the maximum excess volume as also seen in the case of heavy large disks.
Note that as $\beta$ increases, the population of hosts 2,3,4 initially depletes more rapidly than the population of hosts 1 increases, which has the effect that the excess volume initially increases from the value produced by the most disordered macrostate, but then turns around and increases.
This effect is clearly observable in Fig.~\ref{fig:f20}, which shows entropy versus excess volume.
It is remarkably similar to  Fig.~\ref{fig:f8} even though the large and small disks act quite differently.

%
\section{Alternative particles}\label{sec:ze-gra}
%
If gravity can be neglected, as is the case when the plane of the channel is oriented horizontally, the analysis requires no more than two species of particles, constructed from the tiles listed in Table~\ref{tab:t1}.
The geometric and energetic specifications of the two compacts are shown in Table~\ref{tab:t7}.
\begin{table}[b]
  \caption{Two species of compact quasi-particles. The motifs are for $\sigma_\mathrm{L}=2$, $\sigma_\mathrm{S}=1.4$, $H=2.5$. The ID lists the overlapping tiles involved. The excess volume is $\Delta V_m$ and the activation emergy $\epsilon_m$.}\label{tab:t7}
\begin{center}
\begin{tabular}{cccccc} \hline\hline
motif & ID & ~~$m$~~ & category & ~~$\Delta V_m$~~ & $\epsilon_m$  \\ \hline \rule[-2mm]{0mm}{8mm}
\begin{tikzpicture} [scale=0.2]
\draw (0.0,0.0) -- (4.57,0.0) -- (4.57,2.5) -- (0.0,2.5) -- (0,0);
\filldraw [fill=gray, draw=black] (0.7,0.7) circle (0.7);
\filldraw [fill=gray, draw=black] (2.37,1.0) circle (1.0);
\filldraw [fill=gray, draw=black] (3.87,1.8) circle (0.7);
\end{tikzpicture}\hspace{2mm}%
\begin{tikzpicture} [scale=0.2]
\draw (0.0,0.0) -- (4.57,0.0) -- (4.57,2.5) -- (0.0,2.5) -- (0,0);
\filldraw [fill=gray, draw=black] (0.7,1.8) circle (0.7);
\filldraw [fill=gray, draw=black] (2.2,1.0) circle (1.0);
\filldraw [fill=gray, draw=black] (3.87,0.7) circle (0.7);
\end{tikzpicture}
& ~~\textsf{3v},  \textsf{w1}~~ & 1 & compact & $V_\mathrm{t}$ & $pV_\mathrm{t}$ 
\\ \rule[-2mm]{0mm}{6mm}
\begin{tikzpicture} [scale=0.2]
\draw (0.0,0.0) -- (4.57,0.0) -- (4.57,2.5) -- (0.0,2.5) -- (0,0);
\filldraw [fill=gray, draw=black] (0.7,1.8) circle (0.7);
\filldraw [fill=gray, draw=black] (2.37,1.5) circle (1.0);
\filldraw [fill=gray, draw=black] (3.87,0.7) circle (0.7);
\end{tikzpicture}\hspace{2mm}%
\begin{tikzpicture} [scale=0.2]
\draw (0.0,0.0) -- (4.57,0.0) -- (4.57,2.5) -- (0.0,2.5) -- (0,0);
\filldraw [fill=gray, draw=black] (0.7,0.7) circle (0.7);
\filldraw [fill=gray, draw=black] (2.2,1.5) circle (1.0);
\filldraw [fill=gray, draw=black] (3.87,1.8) circle (0.7);
\end{tikzpicture}
&  ~~\textsf{25},  \textsf{46}~~ & 2 & compact & $V_\mathrm{t}$ & $pV_\mathrm{t}$
\\ \hline\hline
\end{tabular}
\end{center}
\end{table}
These compacts are activated from the twofold degenerate pseudo-vacuum,
\begin{equation}\label{eq:30} 
\mathrm{pv}=\{\mathsf{vwvw\cdots v,~ 4545\dots4}\}.
\end{equation}
Each compact particle has two distinct motifs.
However, only one or the other motif will fit into a given slot.
Each compact can be directly followed by a compact of the same species or the other species with exactly one motif fitting.

All microstates which can be constructed under these rules have no more than three successive disks touching the same wall and then always with the small disk in the middle, which ensures mechanical stability.
The combinatorial specifications of the alternative particles are compiled in Table~\ref{tab:t8}.

\begin{table}[htb]
  \caption{Capacity constants $A_m$ and statistical interaction coefficients $g_{mm'}$ for the particles from Table~\ref{tab:t7}.}\label{tab:t8}
\begin{center}
\begin{tabular}{c|c} \hline\hline
$m$~~ & ~~$A_m$  \\ \hline
1~~ & ~~$N-1$ \\
2~~ & ~~$N-1$ \\ 
\hline\hline 
\end{tabular} \hspace{5mm}
\begin{tabular}{c|rr} \hline\hline 
$g_{mm'}$~ & ~~$1$ & ~~$2$  \\ \hline 
$1$ & $1$ & $0$ \\ 
$2$ & $1$ & $1$ \\ 
 \hline\hline
\end{tabular}
\end{center}
\end{table}

Solving Eqs.~(\ref{eq:9}) and (\ref{eq:10}) with $\beta$ from (\ref{eq:4}) and substituting the results into (\ref{eq:7}) and (\ref{eq:8}) reproduces the population densities,
\begin{equation}\label{eq:29} 
\bar{N_1}=\bar{N}_2=\frac{1}{2}\bar{V},
\end{equation} 
for the two species of compacts and the identical relation (\ref{eq:20b}) between entropy and excess volume as previously obtained from four hosts and one tag. 

%
\section{Conclusion and outlook}\label{sec:outl}
%
The work reported in this paper employs a methodology not commonly in use for jammed granular matter -- a methodology shown here and earlier \cite{janac1,janac2} to deliver exact results for nontrivial scenarios.
It promises to do the same for more complex scenarios.
The focus here on disks of two sizes and weights has illuminated the workings of the exact analysis for specific situations.

The interplay between geometry, energetics, and combinatorics is encoded in a set of nonlinear equations for the jammed macrostates.
Of the four types of relevant forces in the system under scrutiny here, the steric forces govern the mechanically stable configurations after jamming, whereas the pressure against mobile pistons, the gravitational force, and imposed random agitations govern the spatial configurations prior to jamming. 
The latter three are encoded in dimensionless parameters, by which all jammed macrostates are characterized. 

The volume and entropy of macrostates, determined by their average particle content, become functions of the dimensionless parameters.
The gradual variation of these parameters is akin to a quasistatic process described by variables analogous to thermodynamic functions.
Some of these processes are found to encounter singularities reminiscent of phase transitions.
A distinctive feature of jammed macrostates is that the same volume and entropy can be produced with very different contents of statistically interacting particles. 

Admittedly, these exact results are highly nongeneric in a broader context.
Jamming conditions in general are known to be nonlocal, which puts them out of reach of our methodology as developed thus far.
However, there are clear paths for the further developement of this approach.
One extension of the work reported here, which we intend to tackle next, promises to cover much new ground.
It permits random sequences of large and small disks of different weights. 
This scenario requires a different jamming protocol, where random agitations take place in a wider channel that allows disks to move past each other. 

If we allow, in the jammed state, multiple small disks to be placed in sequence, the more stringent geometric constraints,
\begin{subequations}\label{eq:17} 
\begin{align}\label{eq:17a} 
& \frac{1}{1+\sqrt{3/4}}<\frac{\sigma_\mathrm{S}}{\sigma_\mathrm{L}}\leq 1, 
\\ \label{eq:17b} 
& \frac{\sigma_\mathrm{L}}{\sigma_\mathrm{S}}<\frac{H}{\sigma_\mathrm{S}}<
1+\sqrt{3/4},
\end{align}
\end{subequations}
must be satisfied, replacing the condition (\ref{eq:1}).
The number of distinct tiles that are realized is doubled from 8 in Table~\ref{tab:t1} to 16 in Table~\ref{tab:t5}. 
The six distinct volumes of the 16 tiles are compiled in Table~\ref{tab:t6}

\begin{table}[b]
  \caption{Distinct tiles that constitute jammed microstates of arbitrary disk sequences subject to the constraints (\ref{eq:2}). Rule: \textsf{v} must be followed by \textsf{w} or 2 etc. Icons: $\sigma_\mathrm{L}=2$, $\sigma_\mathrm{S}=1.4$, $H=2.5$. Rule amendments in square-brackets apply in the presence of centrifugal forces.}\label{tab:t5}
\begin{center}
\begin{tabular}{cccc|cccc} \hline\hline
motif & ~ID~ & rule & vol.~ & ~motif~ & ~ID~ & rule & vol.  \\ \hline \rule[-2mm]{0mm}{8mm}
\begin{tikzpicture} [scale=0.2]
\draw (0.0,0.0) -- (3.94,0.0) -- (3.94,2.5) -- (0.0,2.5) -- (0,0);
\filldraw [fill=gray, draw=black] (1.0,1.5) circle (1.0);
\filldraw [fill=gray, draw=black] (2.94,1.0) circle (1.0);
\end{tikzpicture}
& 1 & $2,6,10,14$ & $V_\mathrm{a}$ &   
\begin{tikzpicture} [scale=0.2]
\draw (0.0,0.0) -- (4.0,0.0) -- (4.0,2.5) -- (0.0,2.5) -- (0,0);
\filldraw [fill=gray, draw=black] (1,1.5) circle (1.0);
\filldraw [fill=gray, draw=black] (3.0,1.5) circle (1.0);
\end{tikzpicture}
& 9 & $1,5,[9]$ & $V_\mathrm{d}$  \\ \rule[-2mm]{0mm}{6mm}

\begin{tikzpicture} [scale=0.2]
\draw (0.0,0.0) -- (3.94,0.0) -- (3.94,2.5) -- (0.0,2.5) -- (0,0);
\filldraw [fill=gray, draw=black] (1,1) circle (1.0);
\filldraw [fill=gray, draw=black] (2.94,1.5) circle (1.0);
\end{tikzpicture}
& 2 & $1,5,9,13$ & $V_\mathrm{a}$ &  
\begin{tikzpicture} [scale=0.2]
\draw (0.0,0.0) -- (4.0,0.0) -- (4.0,2.5) -- (0.0,2.5) -- (0,0);
\filldraw [fill=gray, draw=black] (1.0,1.0) circle (1.0);
\filldraw [fill=gray, draw=black] (3.0,1.0) circle (1.0);
\end{tikzpicture}
& 10 & $2,6,[10]$ & $V_\mathrm{d}$  \\ \rule[-2mm]{0mm}{6mm}

\begin{tikzpicture} [scale=0.2]
\draw (0.0,0.0) -- (2.27,0.0) -- (2.27,2.5) -- (0.0,2.5) -- (0,0);
\filldraw [fill=gray, draw=black] (0.7,1.8) circle (0.7);
\filldraw [fill=gray, draw=black] (1.57,0.7) circle (0.7);
\end{tikzpicture}
& 3 & $4,8,12,16$ & $V_\mathrm{b}$ & 
\begin{tikzpicture} [scale=0.2]
\draw (0.0,0.0) -- (2.8,0.0) -- (2.8,2.5) -- (0.0,2.5) -- (0,0);
\filldraw [fill=gray, draw=black] (0.7,1.8) circle (0.7);
\filldraw [fill=gray, draw=black] (2.1,1.8) circle (0.7);
\end{tikzpicture}
& 11 & $3,7,15,[11]$ & $V_\mathrm{e}$  \\ \rule[-2mm]{0mm}{6mm}

\begin{tikzpicture} [scale=0.2]
\draw (0.0,0.0) -- (2.27,0.0) -- (2.27,2.5) -- (0.0,2.5) -- (0,0);
\filldraw [fill=gray, draw=black] (0.7,0.7) circle (0.7);
\filldraw [fill=gray, draw=black] (1.57,1.8) circle (0.7);
\end{tikzpicture}
& 4 & $3,7,11,15$ & $V_\mathrm{b}$ & 
\begin{tikzpicture} [scale=0.2]
\draw (0.0,0.0) -- (2.8,0.0) -- (2.8,2.5) -- (0.0,2.5) -- (0,0);
\filldraw [fill=gray, draw=black] (0.7,0.7) circle (0.7);
\filldraw [fill=gray, draw=black] (2.1,0.7) circle (0.7);
\end{tikzpicture}
& 12 & $4,8,16,[12]$ & $V_\mathrm{e}$ \\ \rule[-2mm]{0mm}{6mm}

\begin{tikzpicture} [scale=0.2]
\draw (0.0,0.0) -- (3.2,0.0) -- (3.2,2.5) -- (0.0,2.5) -- (0,0);
\filldraw [fill=gray, draw=black] (1,1.5) circle (1.0);
\filldraw [fill=gray, draw=black] (2.5,0.7) circle (0.7);
\end{tikzpicture}
& 5 & $4,8,12,16$ & $V_\mathrm{c}$ &   
\begin{tikzpicture} [scale=0.2]
\draw (0.0,0.0) -- (3.37,0.0) -- (3.37,2.5) -- (0.0,2.5) -- (0,0);
\filldraw [fill=gray, draw=black] (1.0,1.5) circle (1.0);
\filldraw [fill=gray, draw=black] (2.67,1.8) circle (0.7);
\end{tikzpicture}
& 13 & $3,7,15,[11]$ & $V_\mathrm{f}$  \\ \rule[-2mm]{0mm}{6mm}

\begin{tikzpicture} [scale=0.2]
\draw (0.0,0.0) -- (3.2,0.0) -- (3.2,2.5) -- (0.0,2.5) -- (0,0);
\filldraw [fill=gray, draw=black] (1,1) circle (1.0);
\filldraw [fill=gray, draw=black] (2.5,1.8) circle (0.7);
\end{tikzpicture}
& 6 & $3,7,11,15$ & $V_\mathrm{c}$ &  
\begin{tikzpicture} [scale=0.2]
\draw (0.0,0.0) -- (3.37,0.0) -- (3.37,2.5) -- (0.0,2.5) -- (0,0);
\filldraw [fill=gray, draw=black] (1.0,1.0) circle (1.0);
\filldraw [fill=gray, draw=black] (2.677,0.7) circle (0.7);
\end{tikzpicture}
& 14 & $4,8,16,[12]$ & $V_\mathrm{f}$  \\ \rule[-2mm]{0mm}{6mm}

\begin{tikzpicture} [scale=0.2]
\draw (0.0,0.0) -- (3.2,0.0) -- (3.2,2.5) -- (0.0,2.5) -- (0,0);
\filldraw [fill=gray, draw=black] (0.7,1.8) circle (0.7);
\filldraw [fill=gray, draw=black] (2.2,1.0) circle (1.0);
\end{tikzpicture}
& 7 & $2,6,10,14$ & $V_\mathrm{c}$ & 
\begin{tikzpicture} [scale=0.2]
\draw (0.0,0.0) -- (3.37,0.0) -- (3.37,2.5) -- (0.0,2.5) -- (0,0);
\filldraw [fill=gray, draw=black] (0.7,1.8) circle (0.7);
\filldraw [fill=gray, draw=black] (2.37,1.5) circle (1.0);
\end{tikzpicture}
& 15 & $1,5$& $V_\mathrm{f}$  \\ \rule[-2mm]{0mm}{6mm}

\begin{tikzpicture} [scale=0.2]
\draw (0.0,0.0) -- (3.2,0.0) -- (3.2,2.5) -- (0.0,2.5) -- (0,0);
\filldraw [fill=gray, draw=black] (0.7,0.7) circle (0.7);
\filldraw [fill=gray, draw=black] (2.2,1.5) circle (1.0);
\end{tikzpicture}
& 8 & $1,5,9,13$ & $V_\mathrm{c}$ & 
\begin{tikzpicture} [scale=0.2]
\draw (0.0,0.0) -- (3.37,0.0) -- (3.37,2.5) -- (0.0,2.5) -- (0,0);
\filldraw [fill=gray, draw=black] (0.7,0.7) circle (0.7);
\filldraw [fill=gray, draw=black] (2.37,1.0) circle (1.0);
\end{tikzpicture}
& 16 & $2,6$ & $V_\mathrm{f}$ 
\\ \hline\hline
\end{tabular}
\end{center}
\end{table}

\begin{table}[htb]
  \caption{Volume of tiles assuming that the channel has unit cross sections. The numerical values pertain to the icons: $\sigma_\mathrm{L}=2$, $\sigma_\mathrm{S}=1.4$, $H=2.5$.}\label{tab:t6}
\begin{center}
\begin{tabular}{l|l|l} \hline\hline
 & ~vol. & ~num.  \\ \hline \rule[-2mm]{0mm}{6mm}
$V_\mathrm{a}$~ & ~$\sigma_\mathrm{L}+\sqrt{H(2\sigma_\mathrm{L}-H)}$ & ~3.963 
\\ \rule[-2mm]{0mm}{6mm}
$V_\mathrm{b}$~ & ~$\sigma_\mathrm{S}+\sqrt{H(2\sigma_\mathrm{S}-H)}$ & ~2.266 
\\ \rule[-2mm]{0mm}{6mm}
$V_\mathrm{c}$~ & ~$\frac{1}{2}(\sigma_\mathrm{L}+\sigma_\mathrm{S})
+\sqrt{H(\sigma_\mathrm{L}+\sigma_\mathrm{S}-H)}$~~ & ~3.2 
\\ \rule[-2mm]{0mm}{6mm}
$V_\mathrm{d}$~ & ~$2\sigma_\mathrm{L}$ & ~4.0 
\\ \rule[-2mm]{0mm}{6mm}
$V_\mathrm{e}$~ & ~$2\sigma_\mathrm{S}$ & ~2.8 
\\ \rule[-2mm]{0mm}{6mm}
$V_\mathrm{f}$~ & ~$\frac{1}{2}(\sigma_\mathrm{L}+\sigma_\mathrm{S})
+\sqrt{\sigma_\mathrm{L}\sigma_\mathrm{S}}$ & ~3.373 
\\ \hline\hline
\end{tabular}
\end{center}
\end{table}

Constructing a set of statistically interacting particles from this set of 16 tiles with the given successor rules that ensure mechanical stability under jamming is challenging task with no guarantee of success, in general.
However, in this instance we can take advantage of a mapping between this system of jammed disk and a spin-$\frac{3}{2}$ Ising chain, which was previously investigated by the same method \cite{pichs} and shown to represent a system of statistically interacting particles from a set of 17 species.

The mapping is exact if we relax the successor rules in Table~\ref{tab:t5} to include the entries in square brackets. 
This is justified if we consider channels that rotate about their axis and thus produce centrifugal forces on the disks \cite{note2}.
The mapping also requires a switch from an open system (grandcanonical ensemble) realized in the Ising context to a closed system (canonical ensemble) in the jammed-disk context, which requires the introduction of what in a thermodynamic system are chemical potentials. 
The ground is thus cleared for the exact analysis of a system of jammed granular matter with a random mix of grains of two different sizes.
At the same time, a new path opens up for the study of particle segregation \cite{DR97, ESR05, Gray18} with this methodology.

\vfill

\pagebreak


\begin{thebibliography}{100}
\bibitem{note1}
The zero gravity limit can be implemented by tilting the plane of the channel from vertical toward horizontal.
\bibitem{note2}
The relaxed successor rules in Table~\ref{tab:t5} can be eliminated in two ways: (i) by freezing out specific particle species (lifting their activation energies to infinity), (ii) by converting tag particles into cap particles \cite{pichs}.
\end{thebibliography}
\end{document}